\documentclass[twocolumn,prl, aps,showpacs,superscriptaddress,preprintnumbers,amsmath,amssymb]{revtex4-1}
\usepackage{epsfig}
\usepackage{bm}

\usepackage{hyperref}
\hypersetup{backref=true,pagebackref=true,hyperindex=true,colorlinks=true,citecolor=blue, breaklinks=true,urlcolor=
blue,linkcolor=blue,bookmarks=true,bookmarksopen=true}
\usepackage{graphicx}
\usepackage{animate}
\usepackage[]{units}
\usepackage[utf8]{inputenc}
\usepackage[squaren]{SIunits}
\usepackage{upgreek}

\begin{document}

\title{Enhanced Non-Adiabaticity in Vortex Cores due to the Emergent Hall Effect}
\date{\today}

\author{Andr\'e Bisig}
\email{andre.bisig@gmail.com}
\affiliation{Department of Physics, University of Konstanz, 78457 Konstanz, Germany}
\affiliation{Max Planck Institute for Intelligent Systems, 70569 Stuttgart, Germany}
\affiliation{Institute of Condensed Matter Physics, \'Ecole Polytechnique F\'ed\'erale de Lausanne, 1015 Lausanne, Switzerland}
\affiliation{Paul Scherrer Institute, 5232 Villigen PSI, Switzerland}
\affiliation{Institut of Physics, Johannes Gutenberg University Mainz, 55099 Mainz, Germany}

\author{Collins Ashu Akosa}
\affiliation{King Abdullah University of Science and Technology (KAUST), Physical Science and Engineering Division, Thuwal 23955-6900, Saudi Arabia}

\author{Jung-Hwan Moon}
\affiliation{Department of Materials Science and Engineering, Korea University, Seoul 136-713, Korea}

\author{Jan Rhensius}
\affiliation{Department of Physics, University of Konstanz, 78457 Konstanz, Germany}
\affiliation{Paul Scherrer Institute, 5232 Villigen PSI, Switzerland}

\author{Christoforos Moutafis}
\affiliation{Department of Physics, University of Konstanz, 78457 Konstanz, Germany}
\affiliation{Institute of Condensed Matter Physics, \'Ecole Polytechnique F\'ed\'erale de Lausanne, 1015 Lausanne, Switzerland}
\affiliation{Paul Scherrer Institute, 5232 Villigen PSI, Switzerland}

\author{Arndt von Bieren}
\affiliation{Department of Physics, University of Konstanz, 78457 Konstanz, Germany}
\affiliation{Institute of Condensed Matter Physics, \'Ecole Polytechnique F\'ed\'erale de Lausanne, 1015 Lausanne, Switzerland}

\author{Jakoba Heidler}
\affiliation{Department of Physics, University of Konstanz, 78457 Konstanz, Germany}
\affiliation{Institute of Condensed Matter Physics, \'Ecole Polytechnique F\'ed\'erale de Lausanne, 1015 Lausanne, Switzerland}
\affiliation{Paul Scherrer Institute, 5232 Villigen PSI, Switzerland}

\author{Gillian Kiliani}
\affiliation{Department of Physics, University of Konstanz, 78457 Konstanz, Germany}

\author{Matthias Kammerer}
\affiliation{Max Planck Institute for Intelligent Systems, 70569 Stuttgart, Germany}

\author{Michael Curcic}
\affiliation{Max Planck Institute for Intelligent Systems, 70569 Stuttgart, Germany}

\author{Markus Weigand}
\affiliation{Max Planck Institute for Intelligent Systems, 70569 Stuttgart, Germany}

\author{Tolek Tyliszczak}
\affiliation{Advanced Light Source, LBNL, CA 94720 Berkeley, USA}

\author{Bartel Van Waeyenberge}
\affiliation{Department of Solid State Sciences, Ghent University, 9000 Ghent, Belgium}

\author{Hermann Stoll}
\affiliation{Max Planck Institute for Intelligent Systems, 70569 Stuttgart, Germany}

\author{Gisela Sch\"utz}
\affiliation{Max Planck Institute for Intelligent Systems, 70569 Stuttgart, Germany}

\author{Kyung-Jin Lee}
\affiliation{Department of Materials Science and Engineering, Korea University, Seoul 136-713, Korea}
\affiliation{KU-KIST Graduate School of Converging Science and Technology, Korea University, Seoul 136-713, Korea}

\author{Aurelien Manchon}
\email{aurelien.manchon@kaust.edu.sa}
\affiliation{King Abdullah University of Science and Technology (KAUST), Physical Science and Engineering Division, Thuwal 23955-6900, Saudi Arabia}

\author{Mathias Kl\"aui}
\email{klaeui@uni-mainz.de}
\affiliation{Department of Physics, University of Konstanz, 78457 Konstanz, Germany}
\affiliation{Institute of Condensed Matter Physics, \'Ecole Polytechnique F\'ed\'erale de Lausanne, 1015 Lausanne, Switzerland}
\affiliation{Paul Scherrer Institute, 5232 Villigen PSI, Switzerland}
\affiliation{Institut of Physics, Johannes Gutenberg University Mainz, 55099 Mainz, Germany}




\begin{abstract}
We present a combined theoretical and experimental study, investigating the origin of the enhanced non-adiabaticity of magnetic vortex cores. Scanning transmission X-ray microscopy is used to image the vortex core gyration dynamically to measure the non-adiabaticity with high precision, including a high confidence upper bound. Using both numerical computations and analytical derivations, we show that the large non-adiabaticity parameter observed experimentally can be explained by the presence of local spin currents arising from a texture-induced emergent Hall effect. This enhanced non-adiabaticity is only present in two- and three-dimensional magnetic textures such as vortices and skyrmions and absent in one-dimensional domain walls, in agreement with experimental observations.
\end{abstract}

\maketitle


%

The electrical control of magnetic textures through spin angular momentum transfer has attracted a massive amount of interest in the past ten years \cite{Boulle2011}. As spin torque-induced magnetization manipulation exhibits favorable scaling \cite{yamaguchi_2004,yamada_2007}, it underlies novel concepts to store information in non-volatile devices, such as the race-track memory \cite{parkin_2008} or the spin-transfer torque random access memory \cite{kent2015}. Recent progress includes the manipulation of two and three-dimensional chiral magnetic textures, known as magnetic skyrmions \cite{Jonietz2010,Woo2015}, which constitute an inspiring paradigm for potential applications \cite{Fert2013}. The dynamics of a magnetic texture $\mathbf{M}(\mathbf{r},t) = M_s {\bf m}(\mathbf{r},t)$, with M$_s$ being the saturation magnetization, induced by spin transfer torque is usually modeled by the extended phenomenological Landau-Lifshitz-Gilbert (LLG) equation \cite{zhang_2004,thiaville_2005}
\begin{align}
\dot{{\bf m}} =& -\gamma{\bf m}\times\mathbf{H}_\text{eff}+\alpha{\bf m}\times\dot{{\bf m}}  \nonumber\\
&-b_{\rm J}\left(\mathbf{u}\cdot\nabla\right){\bf m}+\beta b_{\rm J}{\bf m}\times\left(\mathbf{u}\cdot\nabla\right){\bf m},
\end{align}
where the first two terms describe the damped precession of magnetization around the effective magnetic field $\mathbf{H}_\text{eff}$, with $\gamma$ denoting the gyromagnetic ratio and $\alpha$ being the viscous Gilbert damping parameter. In the present work, $\alpha$ refers to the damping of the homogeneous magnetic texture, which is different from the effective damping $\alpha_{\rm eff}$ felt by the vortex core, as discussed below. The third term describes the adiabatic momentum transfer from the spin polarized conduction electrons to the local magnetization \cite{tatara_2004}, where $b_{\rm J}{\bf u}=\mathbf{j}_{e}P\mu_\text{B}/e M_\text{S}$ is the spin drift velocity and $P$ is the spin polarization of the conduction electrons. The last term ($\sim \beta b_{\rm J}$) is the so-called non-adiabatic spin transfer torque, which describes the (possibly non-local) torques that do not result from the adiabatic spin transfer \cite{zhang_2004,thiaville_2005}. The magnitude of the non-adiabaticity parameter $\beta$ and in particular the ratio $\beta/\alpha$ determine the efficiency of the spin transfer torque for current-induced domain wall motion, as it governs for instance the domain wall velocity and thus plays a crucial role in the device performance \cite{zhang_2004,thiaville_2005}. However, the physical origin and magnitude of the non-adiabaticity parameter are still under debate and an in-depth understanding of spin transport in magnetic textures is necessary to achieve efficient electrical control of the magnetization.\par

It has been experimentally shown that the ratio $\beta/\alpha$ depends on the domain wall structure, transverse or vortex domain walls in soft magnetic nanostructures \cite{klaui_2008, heyne_2010, eltschka_2010,pollard_2012}, or $180^\circ$-Bloch or N\'eel domain walls in materials with perpendicular magnetic anisotropy \cite{burrowes_2009}. Namely, vortex walls and vortex cores in discs and rectangular elements exhibit a large non-adiabaticity $\beta\approx8-10\alpha$ \cite{heyne_2010,eltschka_2010,pollard_2012} compared to transverse domain walls $\beta\approx\alpha$ \cite{eltschka_2010}, albeit with some uncertainty. This large non-adiabaticity is usually attributed to mistrack between the itinerant spin momentum and the local magnetization \cite{tatara_2004} due to the large texture gradients present in the vortex core (radius of the vortex core $< 10~{\rm nm}$). On the other hand, the non-adiabaticity in very narrow Bloch domain walls (domain wall width of about $1~\text{nm}$) in FePt nanowires is not significantly increased \cite{burrowes_2009}, suggesting that spin mistracking might not be the dominant mechanism for non-adiabaticity.\par

In this Letter, we present a combined theoretical and experimental effort to uncover the origin of non-adiabaticity in magnetic vortex cores. Using scanning transmission X-ray microscopy to image the dynamics of a magnetic vortex core, we first measure the non-adiabaticity parameter with high precision $\beta_{\rm vc} = 0.061 \pm 0.006$ and deduce a high confidence upper bound $\beta_{\rm vc} \le 0.11 \pm 0.01$. Then, based on analytical and numerical considerations, we explain such an enhanced non-adiabaticity by the emergence of a local Hall effect due to the magnetic texture, an effect absent in one-dimensional domain walls.

\begin{figure}[t]
  \begin{center}
  \leavevmode
  \includegraphics[width=0.35\textwidth]{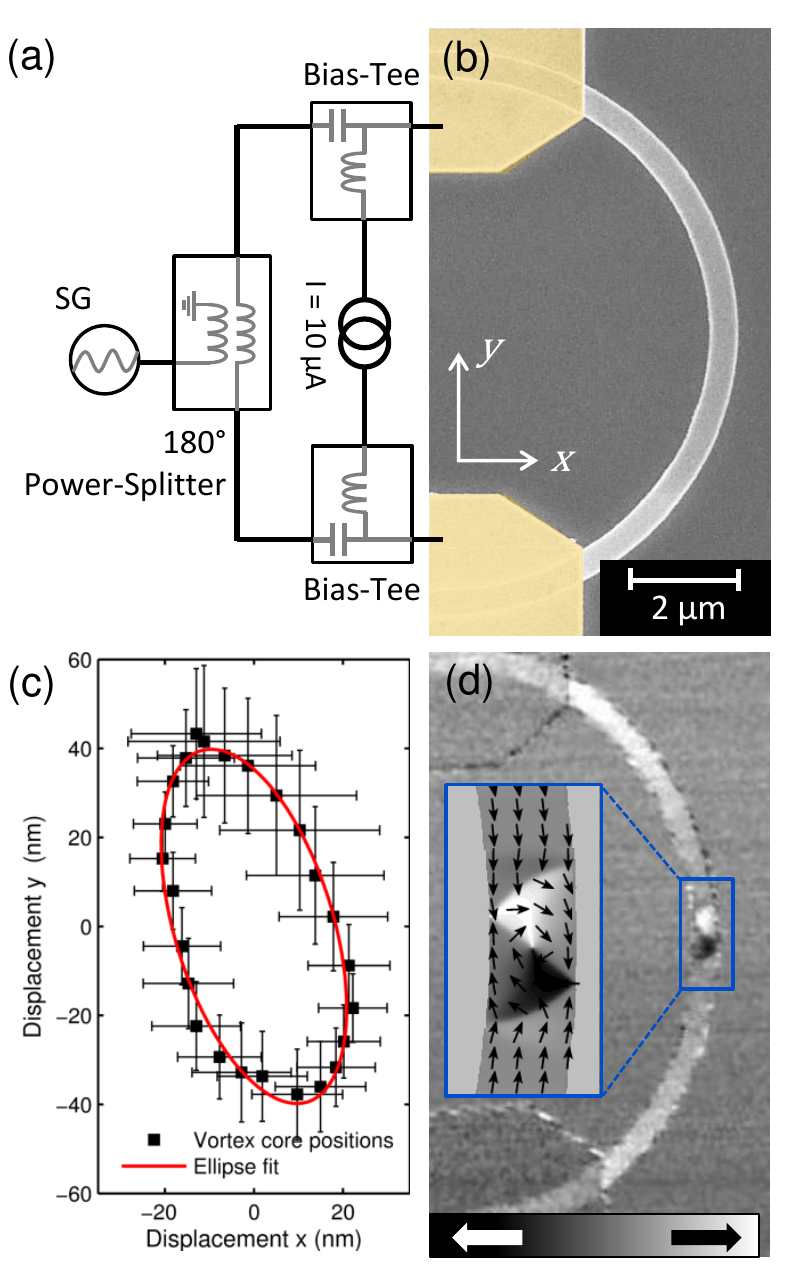}
  \end{center}
  \vspace*{-5mm}
	\caption{(Color online) (a) Schematic illustration of the microwave injection setup. (b) A scanning electron micrograph of the permalloy nanowire under investigation and the two gold contacts on top and at the bottom (yellow). (c) The vortex core positions extracted from the time-resolved images are plotted (black squares). The elliptical vortex core trajectory (red ellipse) is fitted through the data points.	(d) A STXM image showing XMCD contrast of a vortex domain wall at the center of the nanowire.}
	\label{fig:geometry}
\end{figure}

To measure the non-adiabaticity $\beta_{\rm vc}$ of the vortex core, we dynamically imaged the steady state gyration of the vortex core within a vortex domain wall, induced by alternating spin-polarized currents. We study vortex domain walls in a $30~{\rm nm}$ thick and $500~{\rm nm}$ wide permalloy ($\text{Ni}_{80}\text{Fe}_{20}$) half-ring with a radius of $\unit{5}{\upmu m}$, see Fig. \ref{fig:geometry}(b). The half-rings were fabricated on top of a $100$~nm thick silicon nitride ($\text{Si}_3\text{N}_4$) membrane by electron-beam lithography, molecular beam evaporation in UHV and lift-off processing. To improve cooling, a $\unit{150}{nm}$ thick aluminum nitride layer was deposited on top of the structures and on the backside of the $\text{Si}_3\text{N}_4$ membrane. The wires are connected by $\text{Cr}(4~\text{nm})/\text{Au}(100~\text{nm})$ contacts, which are placed more than $\unit{4}{\upmu m}$ from the center of the nanowire [see Fig. \ref{fig:geometry}(b)], to minimize in-plane Oersted fields from vertical electrical currents flowing from the contacts into the nanowire  \cite{bisig_2010}. At the position of the vortex wall, the in-plane field component is negligible ($B < 2\upmu T$) \cite{bisig_2010}, and therefore, we can assume that the vortex gyration is purely induced by the spin transfer torque. After saturation with a magnetic field along the horizontal direction, a vortex domain wall is formed at the center of the nanowire, as shown in  Fig. \ref{fig:geometry}(d). Alternating currents $j_e(t)=\cos(2\pi f t)\cdot\unit{8.7\cdot10^{10}}{Am^{-2}}$ are then injected into the nanowire with different frequencies $f$, while measuring the sample resistance with a small direct current $I=\unit{10}{\rm \upmu A}$, to measure the microwave power within the nanowire and keep the current density constant at all frequencies \cite{bedau_2007}, see  Fig. \ref{fig:geometry}(a). The response of the magnetization to the spin currents was imaged employing time-resolved scanning transmission X-ray microscopy (STXM) at the Advanced Light Source in Berkeley, CA, USA (beamline 11.0.2) \cite{kilcoyne_2003} and at the MAXYMUS endstation, Helmholtz Zentrum Berlin, BESSY II, Germany. In-plane magnetic contrast is obtained by tilting the sample by $60^\circ$ with respect to the X-ray beam and by taking advantage of the X-ray magnetic circular dichroism (XMCD) \cite{schutz_1987}. The data is recorded at the Ni $\text{L}_3$ absorption edge ($852.7$~eV). The lateral resolution is $\approx 25$~nm and the limiting temporal resolution is given by the width of the X-ray photon flashes ($<\unit{70}{ps}$). 

The injection of alternating spin-polarized currents through a vortex structure results in the resonant gyrotropic motion of the vortex core. \cite{kasai_2006,bolte_2008} To analyze the acting torques, we use Thiele's model \cite{thiele_1973,huber_1982,thiaville_2005,shibata_2006}, which describes the motion of the vortex core as a quasi-particle in a restoring potential $V(\mathbf{r})$ 
\begin{equation}
\label{eq:thiele}
{\bf F}_{\rm st} + \nabla_{\mathbf{r}}V(\mathbf{r})+\mathbf{G}\times\left[b_{\rm J}\mathbf{u}-\dot{\mathbf{r}} \right] =  {\cal D}\left[\beta_{\rm vc}b_{\rm J}\mathbf{u}-\alpha_{\rm vc}\dot{\mathbf{r}}\right],
\end{equation}
where $\mathbf{G}=-Gp\hat{\mathbf{e}}_{z}$ is the gyrovector, ${\cal D}_{ij}=\delta_{ij}D$  is the diagonal dissipation tensor \cite{guslienko_2006} and $\alpha_{\rm vc}$ is the damping of the vortex core associated with the Gilbert damping parameter. In a vortex domain wall the parabolic restoring potential is asymmetric and tilted with respect to the current flow  $V({\bf r})=\kappa_{x^\prime}\frac{r_{x^\prime}^2}{2} + \kappa_{y^\prime}\frac{r_{y^\prime}^2}{2}$, where $\kappa_{x^\prime}$, $\kappa_{y^\prime}$ are the potential stiffnesses and $\mathbf{r}=(r_{x^\prime},r_{y^\prime})$ is the displacement of the vortex core from its equilibrium position \cite{moriya_2008,buchanan_2006, bisig_2010}. The coordinate system $(x^\prime, y^\prime)$ is tilted by an angle $\phi$ with respect to the nanowire $(x,y)$ and aligns with the parabolic potential (without loss of generality $\kappa_y^\prime< \kappa_x^\prime$). The resulting motion of the vortex core follows an elliptical trajectory \cite{moriya_2008}. 

Equation (\ref{eq:thiele}) describes the motion of the vortex core as a quasi particle in a restoring potential $V({\bf r})$ under the excitation of the force ${\bf F}_{\rm st}=\mathbf{F}_{\rm ad} + \mathbf{F}_{\rm nad}$ from spin-polarized currents that act via the (non-) adiabatic spin-transfer torque on the vortex core \cite{moriya_2008, heyne_2010, heyne_2011}. The direction of this force, given by the angle $\theta$, can be calculated in the quasi-static limit for $\dot{\bf r}=0$. It depends on the strength of the non-adiabaticity $\beta_{\rm vc}$, on the tilt angle $\phi$ and the asymmetry $r=\kappa_{y^\prime}/\kappa_{x^\prime}$ of the parabolic potential
\begin{equation}
\label{eq:forcedirection}
\tan{\theta}=\frac{G\cos{\phi+\beta_{\rm vc} D\sin\phi}}{-G\sin\phi+\beta_{\rm vc} D \cos\phi}\frac{\kappa_{y^\prime}}{\kappa_{x^\prime}}.
\end{equation}
Therefore, when the shape of the restoring potential $V({\bf r})$ and the direction of the force $\theta$ is known, we can calculate the non-adiabaticity of the vortex core $\beta_{\rm vc}$. Both, $r$ and $\phi$ are a priori unknown for the particular vortex domain wall under investigation and must be determined experimentally, in our case by recording the elliptical vortex core trajectory close to resonance at $f = 210~{\rm MHz}$ \cite{bisig_2010}. The positions of the vortex core and the elliptical vortex core trajectory are plotted in  Fig. \ref{fig:geometry}(c). The error bars indicate the uncertainty of the individual vortex core positions from the experimental images. By fitting an elliptical vortex core trajectory we find $\phi=0.29\pm0.02$~rad and $r=0.19\pm0.01$, the error bars include the uncertainty of the resonance frequency. 

\begin{figure}[t]
  \begin{center}
  \leavevmode
  \includegraphics[width=0.35\textwidth]{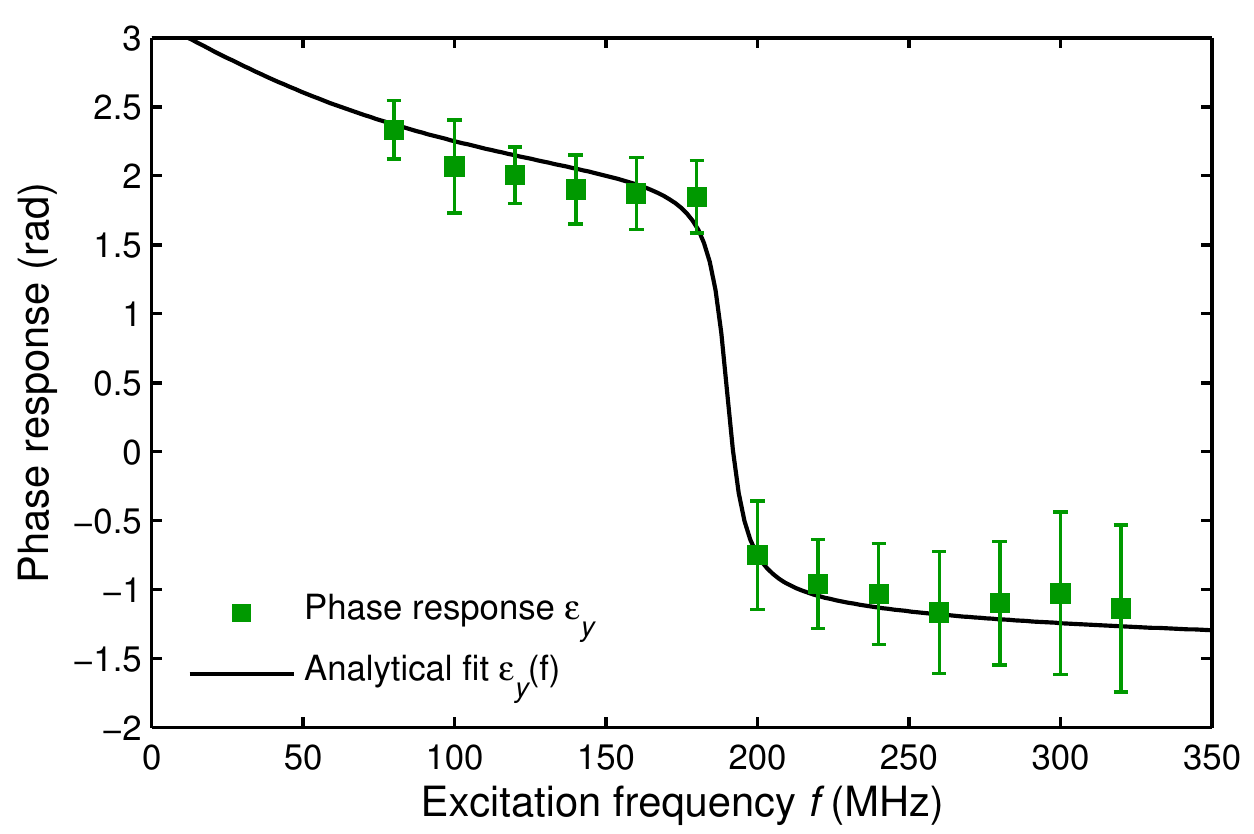}
  \end{center}
  \vspace*{-5mm}
	\caption{(Color online) The phase response of the vortex core gyration is plotted as a function of the excitation frequency $f$, measured experimentally (green squares) and fitted analytically (black line).
		} 
	\label{fig:phaseresponse}
\end{figure}

The phase response $\epsilon_{y}$ of the vortex core to alternating spin-polarized currents (measured along the y-direction), directly depends on the direction $\theta$ of the driving force, and therefore on the non-adiabaticity $\beta_{\rm vc}$, see Fig. \ref{fig:phaseresponse}. Knowing $\phi$ and $r$, we can fit the phase response $\epsilon_y$  with the Thiele model to determine the resonance frequency $f_{\rm r}=\unit{194\pm6}{MHz}$, the non-adiabaticity $\beta_{\rm vc} = 0.061\pm0.006$ and damping $\alpha_{\rm vc}=0.006\pm0.001$. The fit also depends on the ratio between the magnitudes of the gyrovector $\mathbf{G}$ and the dissipation tensor $\cal{D}$, which only moderately depend on the sample geometry \cite{heyne_2010, kruger_2010}. Experimentally, this phase response $\epsilon_y$ was measured by fitting a sinusoidal response through the dynamic differential XMCD contrast at the position of the vortex core. The differential images are obtained by the division of each time-resolved image by the sum of all images. The differential intensity at the region of the vortex core gyration is directly proportional to the displacement of the vortex core in vertical direction.  The error bars include the systematic timing error of the individual time-resolved snapshots given by the electron bunch length and the excitation frequency. 

In addition, equation (\ref{eq:forcedirection}) allows to deduce a maximum bound for the non-adiabaticity from the qualitative behavior of the phase response and through the sign of the denominator by defining a critical non-adiabaticity $\beta_{\rm c}$ when the denominator vanishes
\begin{equation} 
\beta_{\rm c}=\frac{G\sin\phi}{D\cos\phi}.
\end{equation}
The direction $\theta$ of the force $\mathbf{F}_{\rm st}$ is discontinuous and jumps from $-\pi/2$ ($\beta < \beta_{\rm c}$) to $\pi/2$ ($\beta > \beta_{\rm c}$), for fixed tilt angle $\phi$. Experimentally, we observe $\epsilon_y(f\rightarrow0)=\pi$, hence, we can qualitatively determine that $\beta_{\rm vc}<\beta_c = 0.11 \pm 0.01$. Therefore $\beta_{\rm c}(\phi)$ constitutes a high confidence upper limit for the non-adiabaticity that only depends on the angle $\phi$.

This high value for $\beta_{\rm vc}$, and in particular the ratio $\beta_{\rm vc}/\alpha_{\rm vc}=10.4\pm0.3$, is in good agreement with the values obtained by measuring the steady state vortex core displacement \cite{heyne_2010}, by observing thermally assisted domain wall dynamics \cite{eltschka_2010}, or by imaging the frequency dependent vortex core trajectories \cite{pollard_2012}. However, such a high non-adiabaticity at the vortex core is in contrast with the much lower non-adiabaticity measured for one dimensional domain walls \cite{burrowes_2009, eltschka_2010} and to the best of our knowledge, none of the existing models properly account for such a large enhancement. Spin relaxation produces a local non-adiabatic torque $\beta_{\rm sf}\approx0.010\sim \alpha$ \cite{zhang_2004,eltschka_2010} that dominates in smoothly varying magnetic textures \cite{tserkovnyak_2006} but cannot explain the observed non-adiabaticity in vortices. Spin mistracking \cite{tatara_2004} is only significant for extremely sharp domain walls and exponentially vanishes for textures smoother than the spin precession length \cite{tatara_2007,xiao_2006,lee_2013}. For instance, Ref. \onlinecite{lee_2013} estimates $\beta_{\rm sm}\approx 0.0045$ for a domain wall width of $2.7~\text{nm}$. Such spin mistracking-induced non-adiabaticity can be dramatically enhanced in presence of spin-independent disorder \cite{akosa_2015}, but it is questionable whether this effect remains efficient in textures much sharper than the mean free path. Finally, anomalous Hall effect produces non-adiabaticity in vortex cores only \cite{manchon_2011} but in our system this contribution is negligible ($\beta_{\rm AHE}\approx 0.0016$) due to the small Hall angle of permalloy ($\alpha_{\rm H}=1\%$) \cite{nagaosa_2010}. Therefore, it appears that none of the models proposed to date reasonably explain the experimental observations \cite{eltschka_2010,heyne_2010,pollard_2012}.\par

We look for a non-adiabatic torque that is present in vortices only, absent in transverse walls and that does not require extreme magnetization gradients nor strong disorder. Let us consider a clean magnetic system, free from disorder and spin relaxation, with a texture smooth enough so that spin mistracking (i.e. linear momentum transfer) can be neglected. In such a system, the itinerant electron spin experiences an emerging electromagnetic field in the frame of the local quantization axis \cite{Barnes2007}. This field can be expressed in terms of the magnetization spatio-temporal gradient as \cite{Tserkovnyak2008b,Zhang2009b}
\begin{eqnarray}
&&E_i^\sigma=\sigma(\hbar/2e)(\partial_t{\bf m}\times\partial_i{\bf m})\cdot{\bf m},\\
&&B_i^\sigma=-\sigma(\hbar/2e)\epsilon_{ijk}(\partial_j{\bf m}\times\partial_k{\bf m})\cdot{\bf m},
\end{eqnarray}
where $\sigma=\pm$ refers to the spin projection on the local quantization axis ${\bf m}$, $\epsilon_{ijk}$ is Levi-Civita's symbol and $\{i,j,k\}=\{x,y,z\}$. This local electromagnetic field acts oppositely on the two opposite spins and emerges in the presence of magnetization gradient. As a result, the spin-dependent charge current reads
\begin{eqnarray}
&&j_e^{\sigma}=G_\sigma{\bf E}+G_\sigma{\bf E}^s+(G_{\sigma}^2/en){\bf E}\times{\bf B}^\sigma,
\end{eqnarray}
where $G_\sigma$ is the conductivity of spin $\sigma$, $n$ is the electron density and ${\bf E}$ is the applied electric field. The first term is the conventional Ohm's law, the second term is induced by the so-called spin-motive force \cite{Barnes2007}, while the last term is the Hall effect generated by the local magnetic field. This emerging Hall effect is responsible for the topological Hall effect observed in topologically non-trivial magnetic textures such as skyrmions \cite{Neubauer2009}. The induced spin current tensor can be then written
\begin{eqnarray}
{\cal J}^s&=&- b_{\rm J}{\bf m}\otimes {\bf u}+\eta{\bf m}\cdot(\partial_t{\bf m}\times\partial_i{\bf m}){\bf e}_i\nonumber\\
&&+b_{\rm J}\lambda^2[{\bf m}\cdot(\partial_x{\bf m}\times\partial_y{\bf m})]{\bf m}\otimes{\bf m}\times{\bf u},
\end{eqnarray}
where we defined $\eta=g\hbar\mu_{\rm B}G_0/4e^2M_s$, $\lambda^2=\hbar G_0/e^2nP$ and $G_0=G_\uparrow+G_\downarrow$ is the conductivity. The absorption of this spin current produces a torque on the texture, such that ${\bm\tau}=-{\bm\nabla}\cdot{\cal J}^s$, which reads
\begin{eqnarray}\label{eq:tau}
{\bm \tau}&=&b_{\rm J}({\bm\nabla}\cdot{\bf u}){\bf m}-\eta\sum_i[{\bf m}\cdot(\partial_i{\bf m}\times\partial_t{\bf m})]\partial_i{\bf m}\nonumber\\
&&-\lambda^2b_{\rm J}[{\bf m}\cdot(\partial_x{\bf m}\times\partial_y{\bf m})]([{\bf m}\times{\bf u}]\cdot{\bm\nabla}){\bf m}.
\end{eqnarray}
The first term is the conventional adiabatic torque. The second term, proportional to the temporal gradient of the magnetization ($\sim \partial_t{\bf m}$) is a correction to the magnetic damping \cite{Zhang2009b} and the third term is the contribution from the emerging Hall effect.\par


\begin{figure}[t]
  \begin{center}
  \includegraphics[width=8.5cm]{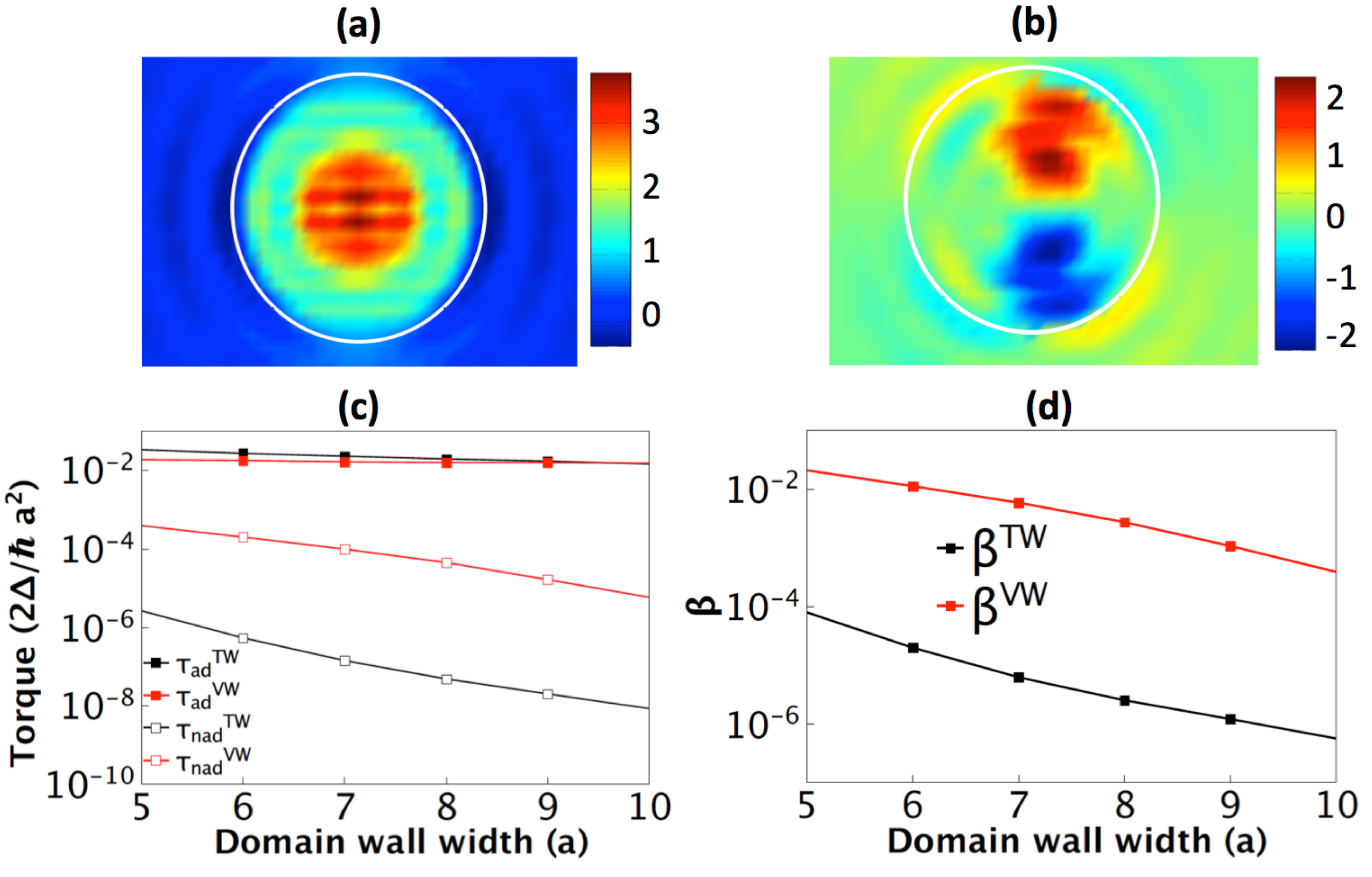}
    \end{center}
  \vspace*{-5mm}
	\caption{(Color online) (a,b) Two dimensional profile of the adiabatic (a) and non-adiabatic torques (b) at the vortex core. The position of the vortex core is indicated by the white solid line and the torque magnitude is expressed in $\tau_{\|,\bot}(\hbar a^2/2\Delta)(\sin^2\vartheta/r_0^2)\times 10^{3}$, $a$ being the lattice parameter. (c,d) Effective adiabatic and non-adiabatic torques (c) and corresponding ratio (d) for a magnetic vortex core (red symbols) and transverse N\'eel wall (black symbol) as a function of the core size and domain wall width, respectively.}
	\label{fig:beta}
\end{figure}

To evaluate the effect of these torques on the magnetic vortex dynamics, we consider an isolated vortex core defined by ${\bf m}=(\sin\vartheta\cos\varphi,\sin\vartheta\sin\varphi,\cos\vartheta)$, with $\vartheta(x,y)=2\tan^{-1}(r/r_0)$ for $r=\sqrt{x^2+y^2}\leq r_0$, $\theta=\pi/2$ for $r_0\leq r\leq R$, and $\phi={\rm Arg}(x,y) +\pi/2$, where $r_0$ ($R$) is the inner (outer) radius of the vortex core. Assuming rigid vortex core motion \cite{thiele_1973}, where $\partial_t{\bf m}=-({\bf v}\cdot{\bm\nabla}){\bf m}$, we obtain the velocity ${\bf v}=v_x{\bf x}+v_y{\bf y}$ of the vortex core
\begin{eqnarray}
{\cal C}\alpha_{\rm eff}v_x-v_y&=&\beta_{\rm eff}b_{\rm J},\\
v_x+{\cal C}\alpha_{\rm eff}v_y&=&b_{\rm J},
\end{eqnarray}
where ${\cal C}=1+\ln\sqrt{R/r_0}$, $\alpha_{\rm eff}=\alpha+(7/3{\cal C})(\eta/r_0^2)$ and $\beta_{\rm eff}=\beta+(17/12)(\lambda^2/r_0^2)$ are the renormalized damping and non-adiabatic coefficient. Here, $\beta\approx\alpha$ is the constant non-adiabaticity parameter measured, e.g., in transverse walls. Using $G_0= 10^7$ $\Omega^{-1}\cdot$m$^{-1}$, $n=10^{29}$ m$^{-3}$, $P=0.5$, and $M_s=800$ emu/cc, we get $\eta=0.24$ nm$^2$ and $\lambda^2=0.8$ nm$^2$. As a consequence, we obtain $\beta_{\rm eff}\approx$0.055 and $\alpha_{\rm eff}\approx$0.017. These estimations, derived in the framework of the $s$-$d$ model, disregard the $spd$ hybridization usually encountered in transition metals. Furthermore, they assume adiabatic spin transport, neglecting spin mistracking and thereby underestimating the non-adiabaticity. Nevertheless, it clearly indicates that the local spin current induced by the emergent Hall effect dramatically enhances the non-adiabaticity of the spin-texture, an effect absent in one-dimensional domain walls, where only the magnetic damping is enhanced. Indeed, this local Hall effect involves a two-dimensional derivative [$\sim \partial_x{\bf m}\times\partial_y{\bf m}$ in Eq. (\ref{eq:tau})], which vanishes for gradients along one dimension only.


To obtain a more accurate estimate of the non-adiabaticity of the vortex core, we numerically compute the spin transport in a vortex core using the tight-binding model described in Ref. \onlinecite{akosa_2015}. The torque is obtained from the local nonequilibrium spin density $\delta{\bf S}$, such that ${\bm\tau}=(2\Delta/\hbar){\bf m}\times\delta{\bf S}$, $\Delta$ being the exchange parameter. It is then parsed into adiabatic and non-adiabatic components, ${\bm\tau}=\tau_{\rm ad}\nabla_x{\bf m}+\tau_{\rm nad}{\bf m}\times\nabla_x{\bf m}$, reported on Fig. \ref{fig:beta}(a) and (b), respectively. While $\tau_{\rm ad}$ is distributed homogeneously around the center of the core, $\tau_{\rm nad}$ is asymmetric along the direction transverse to the applied electric field, reflecting the Hall effect origin of the torque. To evaluate the effective non-adiabaticity parameter, the torque components must be averaged over the volume $\Omega$ of the core texture according to Thiele's equation 
\begin{eqnarray}
\langle\tau_{\rm ad}\rangle=&\frac{\int d\Omega \tau_{\rm ad}(\partial_x\vartheta\partial_y\varphi-\partial_y\vartheta\partial_x\varphi)\sin\vartheta}{\int d\Omega(\partial_x\vartheta\partial_y\varphi-\partial_y\vartheta\partial_x\varphi)\sin\vartheta},\\
\langle\tau_{\rm nad}\rangle=&\frac{\int d\Omega \tau_{\rm nad}[(\partial_x\vartheta)^2+\sin\vartheta(\partial_x\varphi)^2]}{\int d\Omega [(\partial_x\vartheta)^2+\sin\vartheta(\partial_x\varphi)^2]}.
\end{eqnarray}
The results are shown in Fig. \ref{fig:beta}(c) as a function of the width of the core (red symbols). To compare, we also inserted the values obtained in the case of a transverse N\'eel domain wall of same width (black symbols). While the adiabatic torques are about the same order, the non-adiabaticity in the vortex core is much larger than in the transverse wall, which results in a large non-adiabaticity ratio [see Fig. \ref{fig:beta}(d)]. While only spin mistracking is present for the transverse wall, in vortex cores in addition the emergent Hall effect is present.\par

In conclusion, we have determined the non-adiabaticity locally of the vortex core $\beta_{\rm vc}=0.061\pm0.006$ using a highly sensitive phase shift method. In addition to the known minimum bound of the non-adiabaticity \cite{pollard_2012}, we derived a maximum bound by analyzing the qualitative behaviour of the phase response, in summary we conclude $0.041<\beta_{\rm vc}<0.11$. To explain such a high non-adiabaticity at the vortex core we proposed that the texture-induced emergent Hall effect generates non-local non-adiabatic torques. The values obtained by the theory are consistent with the experimental observations. These results are particularly encouraging for the manipulation of current-driven two and three dimensional textures such as skyrmions.

\begin{acknowledgments}

The authors acknowledge support by the German Science Foundation (DFG SFB 767, KL1811, MAINZ GSC 266), the ERC (MASPIC 2007-Stg 208162), the EU (RTN Spinswitch, MRTN CT-2006-035327 , MAGWIRE FP7-ICT-2009-5 257707), COMATT and the Swiss National Science Foundation. Part of this work was carried out at the MAXYMUS scanning X-ray microscope at HZB, BESSY II in Berlin. The Advanced Light Source is supported by the Director, Office of Science, Office of Basic Energy Sciences, of the U.S. Department of Energy under Contract No. DE-AC02-05CH11231. A.M. and C.A. are supported by the King Abdullah University of Science and Technology (KAUST).

\end{acknowledgments}


%


\end{document}